\documentclass[10pt]{article}

\usepackage[utf8]{inputenc}
\usepackage[T1]{fontenc}
\usepackage[margin=0.85in]{geometry}
\usepackage{booktabs}
\usepackage{tabularx}
\usepackage{array}
\usepackage{amsmath,amssymb}
\usepackage{graphicx}
\usepackage{hyperref}
\usepackage{authblk}
\usepackage{caption}
\usepackage{enumitem}
\usepackage{float}
\usepackage{titlesec}
\usepackage{xcolor}
\usepackage{xurl}

\hypersetup{
  colorlinks=true, linkcolor=blue!50!black, citecolor=blue!50!black, urlcolor=blue!50!black,
  pdftitle={Interaction Density as a Behavioural Signature of Exhibit Type},
  pdfauthor={R A Udaya Rakshith, Inavamsi Enaganti, Umang J Gala}
}

\newcolumntype{L}[1]{>{\raggedright\arraybackslash}p{#1}}
\newcolumntype{C}[1]{>{\centering\arraybackslash}p{#1}}
\newcolumntype{R}[1]{>{\raggedleft\arraybackslash}p{#1}}

\titlespacing*{\section}{0pt}{8pt}{4pt}
\titlespacing*{\subsection}{0pt}{6pt}{3pt}
\title{\large\bfseries Interaction Density as a Behavioural Signature of Exhibit Type:\\
A Minimal-Log Study from a Two-Venue Science Experience Centre}

\author[1]{R A Udaya Rakshith}
\author[1]{Inavamsi Enaganti}
\author[1]{Umang J Gala}
\affil[1]{PAIR Labs, Param Foundation, Bengaluru, India \\ \texttt{pair@paraminnovation.org}}

\date{July 2026}

\begin{document}
\maketitle
\thispagestyle{empty}

\begin{abstract}
\noindent Understanding how visitors engage with interactive exhibits usually calls for either labour-intensive manual observation or invasive multimodal sensing -- eye-tracking, cameras, wearables -- that few science centres can deploy at scale. We ask how much can be learned instead from the handful of fields that most touch-enabled exhibits already log by default: a session's start time, end time, and press count. Analysing 2{,}816 visitor sessions across eight exhibits at two venues of a science experience centre in Bengaluru, India, we derive interaction density -- presses per second -- as a simple behavioural signature, and use it to distinguish fast-paced games from slower, deliberate quizzes. Density does so cleanly (Mann-Whitney $r=0.556$) and predicts exhibit type on its own with a cross-validated $\mathrm{AUC}=0.778$. But the data complicates the obvious story: games are not just more intense, visitors also dwell on them longer ($r=0.172$), reversing the intuitive trade-off between intensity and duration -- traced to exhibits whose escalating difficulty creates open-ended re-engagement loops rather than fixed endpoints. Density is not a universal replacement for existing metrics either: raw press count alone explains far more variance in dwell time ($R^2=0.527$) than density does ($R^2=0.081$), though combining both improves on either alone ($R^2=0.667$). Exhibit-level anomalies, a cross-venue replication check, and a session-length censoring artefact further stress-test rather than simply confirm these results. The broader case we make is methodological: minimal, privacy-preserving interaction logs -- not additional sensors -- can already support rigorous, falsifiable behavioural research at any science centre with touch-enabled exhibits.
\end{abstract}


\section{Introduction}

Interactive exhibits are central to how science centres engage visitors, yet quantifying how different exhibit types produce different visitor behaviour typically requires costly manual observation or invasive sensing. We investigate a minimal alternative: whether three automatically logged values per session -- \emph{start time}, \emph{end time}, and \emph{touch count} -- support meaningful, falsifiable inference about exhibit-type behaviour.

The study was conducted at PARSEC (Param Science Experience Centre), operated by Param Foundation across two venues in Bengaluru, India. PARSEC treats the centre itself as a live research environment: exhibits generate data continuously, and that data feeds back into design decisions. From the three logged values we derive \emph{interaction density}, presses per second within a session, and test three hypotheses relating it, dwell time, and press count to whether an exhibit is a \emph{game} (goal-directed, scored, progressively difficult) or a \emph{quiz} (structured questions, defined responses).

Our contributions: (1) a minimal three-field logging methodology requiring no hardware beyond the exhibit's own touch interface; (2) an empirical test of three pre-registered hypotheses, two of which are contradicted or reversed by the data; (3) documented exhibit-level anomalies, a cross-venue replication check, and a session-length censoring artefact, each of which qualifies the main findings without overturning them.

\section{Related Work}

Digital exhibits create an opportunity to observe visitor behaviour directly through interaction logs, rather than only through on-site human observation. The traditional approach to quantifying visitor attention is manual tracking-and-timing, in which an observer records how long each visitor dwells at an exhibit and how many stops they make across a gallery \cite{serrell1997}. This methodology established dwell time, and its area-normalised variant, as the standard behavioural proxy for visitor engagement in museum evaluation. It is well suited to comparing exhibits at the gallery level, but it does not scale to continuous, per-session monitoring and does not by itself capture how intensely a visitor interacts within a single visit.

Instrumented exhibits allow log-based features to complement dwell time directly. Emerson et al.\ \cite{emerson2020} model visitor dwell time at a tabletop science-museum exhibit using multimodal sensor data, including an interaction-log feature they term \emph{taps per second} -- the total number of screen taps scaled by session duration. This is conceptually identical to the interaction density $\delta$ used in the present study, though in their work it is one predictor among facial-expression, eye-gaze, and posture features derived from a heavily instrumented 86-visitor study, rather than a standalone behavioural signature computed from unaugmented touch logs at the scale of thousands of sessions. More recently, Ferrato et al.\ \cite{ferrato2025} model museum-visitor dwell time from behavioural trace data using cross-validated predictive models, reflecting a broader move in visitor analytics toward held-out evaluation rather than resubstitution accuracy alone -- a concern we address directly for our density-based classifier (H2, Section~\ref{sec:results}).

Our contribution relative to this literature is methodological rather than instrumental: rather than adding data channels (eye tracking, facial expression, posture) to a dwell-time model, we ask how much can be inferred from the two fields that most touch-enabled exhibits already log by default -- timestamps and press counts -- without any additional sensing hardware.

\section{Setting and Exhibits}

PARSEC does not separate research from operation; the centre is the laboratory, and every visitor session is simultaneously an experience and a data point. Eight logged exhibit identifiers were selected because they classify unambiguously into game or quiz type by primary interaction mechanic. Two exhibits, Color Perception and Pace the Clock, are deployed identically at both venues and are logged as separate exhibit identifiers per venue; together with three single-venue game exhibits and three quiz exhibits, this gives five game-type and three quiz-type exhibit identifiers (eight in total), summarised in Table~\ref{tab:exhibits} at the level of six conceptual exhibit designs.

\begin{table}[H]
\centering\small
\begin{tabularx}{\linewidth}{@{}L{2.6cm}C{0.9cm}L{1.6cm}X@{}}
\toprule
\textbf{Exhibit} & \textbf{Type} & \textbf{Venue} & \textbf{Mechanic} \\
\midrule
Color Perception & Game & WF, JN & Identify the odd-shaded tile; grid grows, contrast narrows each level, no fixed endpoint. \\
Circle Drawing & Game & WF & Two-player race to complete circles; the only co-play exhibit in the set. \\
Pace the Clock & Game & WF, JN & Stop a hidden timer at a target time; accuracy window tightens with difficulty. \\
Yogic Nature Discovery & Quiz & WF & Sequential personality quiz; reward (avatar) revealed only at the end. \\
Road Sign Quiz & Quiz & WF & Multiple-choice road-sign knowledge quiz; non-adaptive, one question at a time. \\
Lilavati Ancient & Quiz & WF & Maths puzzles (fractions, algebra, ratios) from the treatise \emph{Lilavati}; density behaves game-like despite quiz classification (Sec.~\ref{sec:results}). \\
\bottomrule
\end{tabularx}
\caption{Exhibits studied, by conceptual design. WF = Whitefield, JN = Jayanagar. Color Perception and Pace the Clock each contribute two exhibit identifiers (one per venue) to the analysis in Table~\ref{tab:pipeline}.}
\label{tab:exhibits}
\end{table}

\section{Hypotheses}

For session $i$ with press count $P_i$ and duration $T_i$ (seconds), interaction density is $\delta_i = P_i/T_i$. We test:
\begin{description}[leftmargin=1.2cm,style=nextline,itemsep=1pt,topsep=2pt]
\item[H1] Games show higher density than quizzes, and quizzes show longer dwell time than games (games intense-but-short; quizzes slow-but-longer).
\item[H2] Density alone classifies exhibit type (game vs.\ quiz) with $\mathrm{AUC}\geq 0.70$.
\item[H3] Density is a stronger predictor of dwell time than raw press count alone.
\end{description}

\section{Dataset and Methods}

Data were collected over 7 June -- 7 July 2026 across both PARSEC venues via automated firmware logging (ESP32/Raspberry Pi), using a 60-second inactivity window for session boundaries and 150\,ms touch debounce. Raw data: 8{,}191 sessions across 26 exhibit identifiers.

Sessions with duration $<10$s were excluded (non-engagement events, threshold set prior to analysis); no session in the raw log has zero presses, so this single filter fully implements our intended exclusion criteria (Table~\ref{tab:pipeline}). Eighteen of the 26 logged exhibit identifiers -- including a device hostname logged in place of an exhibit identifier and a Jayanagar feedback form that is not an exhibit -- fall outside the eight game/quiz identifiers analysed here and are simply not part of the game/quiz subset; no separate exclusion step is needed for them.

\begin{table}[H]
\centering\small
\begin{tabularx}{\linewidth}{@{}L{6.3cm}R{2.8cm}@{}}
\toprule
\textbf{Stage} & \textbf{Sessions} \\
\midrule
Raw & 8{,}191 \\
After duration $\geq$10s filter (presses $\geq$1 already holds for every raw session) & 6{,}261 \\
Game/quiz analysis subset (Game / Quiz) & 2{,}816 (1{,}744 / 1{,}072) \\
\bottomrule
\end{tabularx}
\caption{Cleaning pipeline and final sample.}
\label{tab:pipeline}
\end{table}

Classification (game/quiz) was performed independently by two researchers from exhibit descriptions and primary interaction mechanic, not names; one disagreement was resolved by operational description. Dwell time and density are right-skewed for all groups (Shapiro-Wilk $p<0.001$), so between-group tests use Mann-Whitney $U$, with effect size $r = 1 - 2U/(n_1 n_2)$. H2 uses a logistic classifier on density; we report both the in-sample AUC and a 5-fold stratified cross-validated AUC to check for overfitting. H3 uses OLS regression on $\log(\text{duration\_s})$, and we likewise report 5-fold cross-validated $R^2$ alongside the in-sample fit.

\section{Results}
\label{sec:results}

\begin{table}[H]
\centering\small
\begin{tabularx}{\linewidth}{@{}L{5.4cm}R{2.2cm}R{2.2cm}R{2.4cm}@{}}
\toprule
& \textbf{Games} & \textbf{Quizzes} & \textbf{Effect} \\
\midrule
Median density (p/s) & 0.365 & 0.202 & $r=\mathbf{0.556}$ \\
Median dwell (s) & 170.2 & 117.8 & $r=\mathbf{0.172}$ \\
\bottomrule
\end{tabularx}
\caption{H1: density and dwell time, games vs.\ quizzes ($n=1{,}744$ / $1{,}072$). Both $U$-tests $p<0.0001$ (density: $U=1{,}454{,}899$; dwell: $U=1{,}095{,}508$).}
\label{tab:h1}
\end{table}

\noindent\textbf{H1 density -- supported.} $r=0.556$ (medium-to-large). Games show 52\% higher density than quizzes at the mean, and 81\% higher at the median; interquartile ranges (games 0.294--0.467\,p/s, quizzes 0.139--0.311\,p/s) overlap only slightly, at their adjoining edges (quiz 75th percentile 0.311 vs.\ game 25th percentile 0.294).

\begin{figure}[H]
\centering
\includegraphics[width=0.62\linewidth]{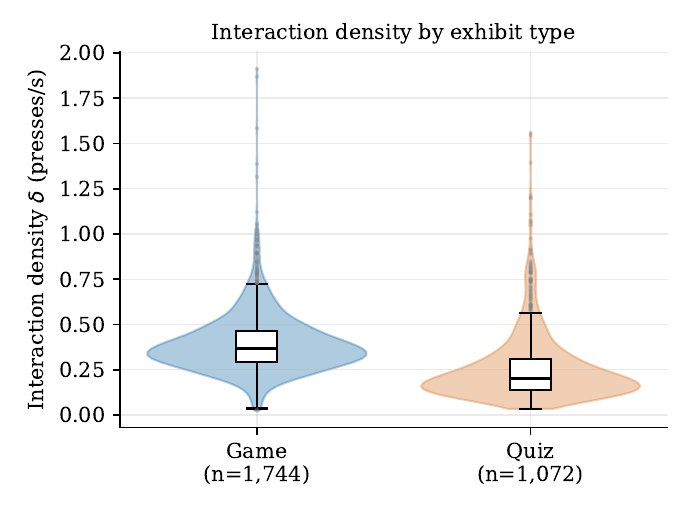}
\caption{Distribution of interaction density by exhibit type. Boxes show the median and interquartile range; violins show the full density.}
\label{fig:density_by_type}
\end{figure}

\noindent\textbf{H1 dwell -- contradicted.} Games dwell 44\% \emph{longer} at the median, not shorter as hypothesised ($r=0.172$, small but a clear directional reversal). Color Perception (median 336--366s at both venues, 75th percentile $>$700s) drives this: its continuously escalating difficulty creates a re-engagement loop with no fixed endpoint. \emph{Revised finding: games here are both more intense and longer, not a trade-off.}

\textbf{Exhibit-level anomaly.} Lilavati Ancient (quiz) shows mean density 0.37\,p/s -- inside the game range -- because solving a maths puzzle requires repeated input (trying, erring, correcting), unlike a single deliberate quiz selection. It also has the lowest median dwell in the set (71s), an ambiguous short-and-dense signature (rapid completion or frustration) that log data alone cannot resolve. Yogic Nature Discovery gives the cleanest quiz signature (mean 0.17\,p/s): one press per prompt, reward deferred to the end.

\begin{figure}[H]
\centering
\includegraphics[width=0.72\linewidth]{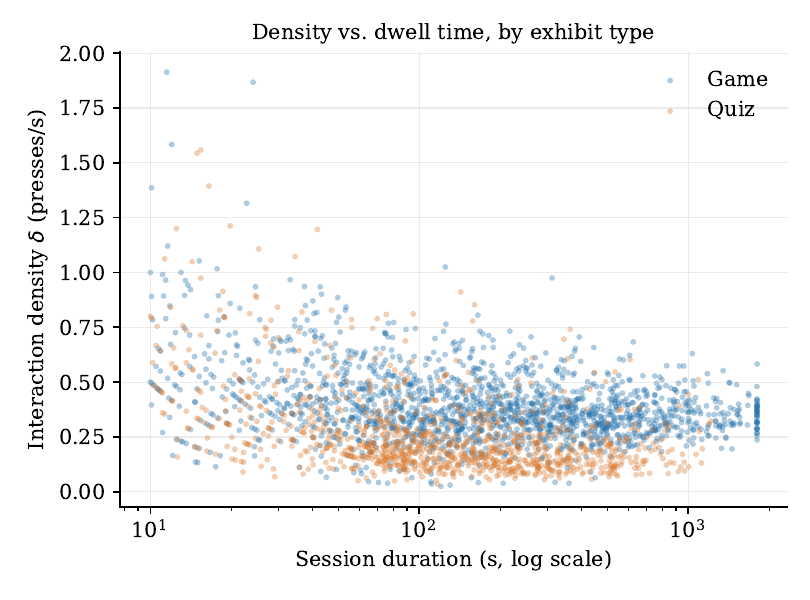}
\caption{Interaction density against session duration (log scale), by exhibit type. The two exhibit-level anomalies sit at opposite ends of the duration axis: Lilavati Ancient (quiz, short-and-dense) and Color Perception (game, long-and-dense).}
\label{fig:density_vs_dwell}
\end{figure}

\begin{table}[H]
\centering\small
\begin{tabularx}{\linewidth}{@{}L{4.7cm}R{2.9cm}X@{}}
\toprule
\textbf{Metric} & \textbf{Value} & \textbf{Interpretation} \\
\midrule
H2: AUC, in-sample & 0.778 & \\
H2: AUC, 5-fold CV & $0.778\pm0.023$ & Matches the in-sample value almost exactly -- \textbf{not overfitting}; exceeds 0.70 threshold, \textbf{supported}. \\
H3: $R^2$, density alone & 0.081 / 0.079 & in-sample / CV \\
H3: $R^2$, presses alone & 0.527 / 0.526 & in-sample / CV \\
H3: $R^2$, both combined & 0.667 / 0.666 & in-sample / CV \\
\bottomrule
\end{tabularx}
\caption{H2 (logistic classifier, density $\to$ exhibit type) and H3 (OLS on $\log$ dwell time). Cross-validated figures are 5-fold means. Press count explains 6.5$\times$ more variance than density alone -- \textbf{reversed} relative to H3. The combined model gains 14\,pp over press count alone, so density still carries independent signal.}
\label{tab:h2h3}
\end{table}

\begin{figure}[H]
\centering
\includegraphics[width=0.55\linewidth]{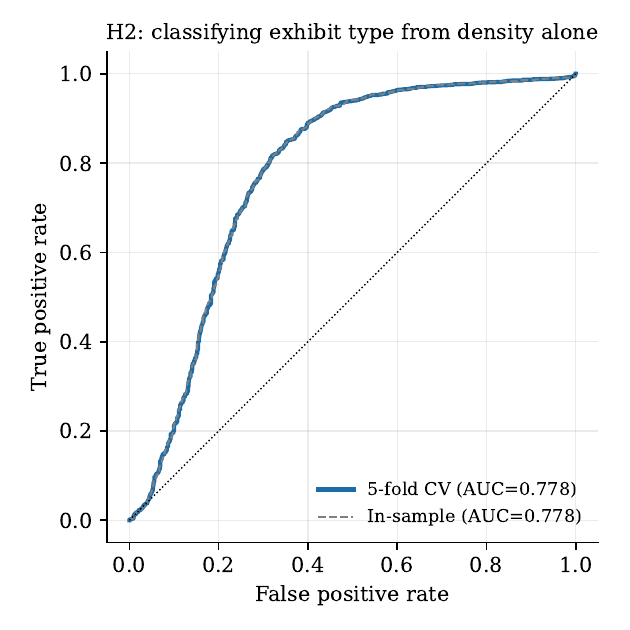}
\caption{ROC curve for the H2 classifier (exhibit type from density alone). The 5-fold cross-validated curve sits almost on top of the in-sample curve, visually confirming that the classifier generalises rather than memorising the training data.}
\label{fig:roc}
\end{figure}

\noindent\textbf{Cross-venue check.} Color Perception's mean density replicates to within roughly 12\% across venues (WF 0.40\,p/s vs.\ JN 0.36\,p/s) and its median dwell to within 9\% (WF 366s vs.\ JN 336s) -- density appears to be substantially a property of the exhibit, not the venue, though the two venues are not identical. Pace the Clock replicates closely on density (0.39 vs.\ 0.38\,p/s) but \emph{not} on dwell (100s WF vs.\ 141s JN, +41\%), suggesting a visitor-population effect independent of interaction pace -- exactly the kind of signal a single dwell-only metric would obscure.

\begin{figure}[H]
\centering
\includegraphics[width=0.85\linewidth]{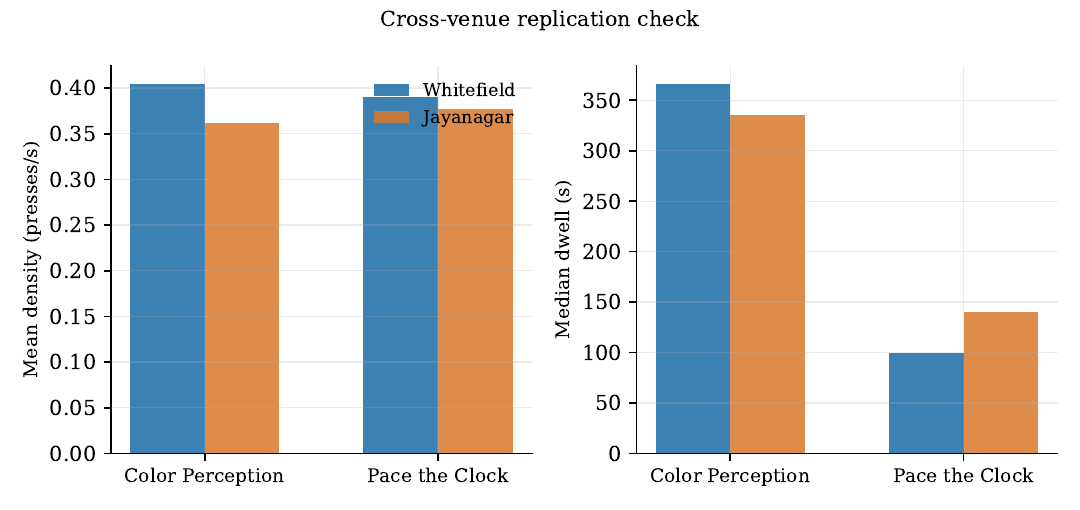}
\caption{Cross-venue replication check for the two exhibits deployed at both Whitefield and Jayanagar.}
\label{fig:cross_venue}
\end{figure}

\textbf{Session-length censoring artefact.} The firmware/session logic appears to enforce an approximate 1{,}800-second (30-minute) session ceiling: 46 of 8{,}191 raw sessions (0.6\%) end within 5 seconds of this value, disproportionately at Color Perception (4.3\% of its Whitefield sessions, 5.2\% at Jayanagar). These are plausibly right-censored engagements -- the visitor may still have been interacting when the session was force-closed -- rather than natural session ends. Excluding them lowers Color Perception's median dwell by 4--9\% (WF: 365.7s $\to$ 351.2s; JN: 335.8s $\to$ 307.0s), but its dwell time remains far longer than any other exhibit in the set, so the H1 dwell reversal reported above is not an artefact of this censoring.

\begin{table}[H]
\centering\small
\begin{tabularx}{\linewidth}{@{}L{1cm}X R{3.3cm}@{}}
\toprule
& \textbf{Hypothesis} & \textbf{Verdict} \\
\midrule
H1a & Density: game $>$ quiz & \textbf{Supported} \\
H1b & Dwell: quiz $>$ game & \textbf{Contradicted} \\
H2 & Density fingerprints type & \textbf{Supported (CV)} \\
H3 & Density $>$ presses (predictor) & \textbf{Reversed} \\
\bottomrule
\end{tabularx}
\caption{Summary of verdicts.}
\end{table}

\section{Discussion and Limitations}

Interaction density robustly distinguishes game from quiz exhibits and supports type classification beyond our pre-specified threshold under cross-validation, consistent with a genuine behavioural difference in physical interaction style rather than an artefact of resubstitution. However, two of three hypotheses were not confirmed as stated: the assumed intensity--duration trade-off is contradicted (driven largely by Color Perception's re-engagement loop, and not attributable to the session-length censoring artefact described above), and density is dominated by press count as a duration predictor, though it retains independent value in combination. We read both reversals as more informative than confirmations would have been: a single scalar metric -- dwell time alone, or density alone -- is insufficient, and a two-dimensional intensity/duration framing better accommodates exhibit-level heterogeneity (e.g., Lilavati's short-and-dense profile vs.\ Color Perception's long-and-dense profile).

\textbf{Limitations.} No visitor identifier is logged; sessions are treated as independent, though group visits could generate correlated sessions undetected by this design. Press semantics vary across exhibits (a tile selection is not a scroll gesture), complicating raw cross-exhibit comparison. The quiz group (3 exhibit identifiers) is smaller than the game group (5), limiting the precision of type-level generalisation. An apparent $\sim$1{,}800-second session ceiling right-censors a small fraction of sessions (0.6\% overall, up to 5.2\% at Color Perception); we show above that this does not change the qualitative ranking of exhibits by dwell time, but the true dwell time of censored sessions is unknown. No demographic, video, or additional sensor data were used; ambiguous cases (e.g., Lilavati's short-and-dense pattern) are hypotheses for on-site confirmation, not established facts.

\section{Conclusion}

Using only three logged fields per session, we tested three pre-specified hypotheses relating exhibit type to visitor interaction at a two-venue science experience centre. Interaction density reliably distinguishes game from quiz exhibits (cross-validated AUC $=0.778\pm0.023$), but the assumed intensity--duration trade-off does not hold, and raw press count -- not density -- dominates as a predictor of dwell time, with the two measures acting as complements rather than substitutes. Together with exhibit-level anomalies, a partial cross-venue replication, and a quantified session-censoring check, these results suggest a minimal three-field log is sufficient to generate falsifiable -- and here, partly falsified -- engagement hypotheses, offering a low-cost template for other science centres with touch-enabled exhibits.

\bigskip
\noindent\textbf{Data and code availability.} The session-level dataset underlying this paper (raw session log, CSV) is included as an ancillary file with this arXiv submission. A curated copy is additionally being submitted to AI Kosh and Harvard Dataverse for long-term archival; those listings are pending review at the time of writing.


\begin{thebibliography}{9}

\bibitem{serrell1997}
Beverly Serrell. 1997. Paying Attention: The Duration and Allocation of Visitors' Time in Museum Exhibitions. \textit{Curator: The Museum Journal} 40, 2 (1997), 108--125. \url{https://doi.org/10.1111/j.2151-6952.1997.tb01292.x}

\bibitem{emerson2020}
Andrew Emerson, Nathan Henderson, Jonathan Rowe, Wookhee Min, Seung Lee, James Minogue, and James Lester. 2020. Investigating Visitor Engagement in Interactive Science Museum Exhibits with Multimodal Bayesian Hierarchical Models. In \textit{Proceedings of the 21st International Conference on Artificial Intelligence in Education (AIED 2020)}, LNCS vol.\ 12163. Springer, 165--176. \url{https://doi.org/10.1007/978-3-030-52237-7_14}

\bibitem{ferrato2025}
Alessio Ferrato, Giuseppe Sansonetti, and Marko Tkal\v{c}i\v{c}. 2025. On the Role of Dwell Time for Implicitly Profiling Museum Visitors. In \textit{Proceedings of the 33rd ACM Conference on User Modeling, Adaptation and Personalization (UMAP '25)}. ACM, New York, NY, USA. \url{https://doi.org/10.1145/3774935.3806782}

\end{thebibliography}
\end{document}